\documentclass[twocolumn,showpacs,preprintnumbers,amsmath,amssymb,aps]{revtex4}
\usepackage{amsmath}
\usepackage{graphicx}
\usepackage{dcolumn}
\newcolumntype{d}[1]{D{.}{.}{#1}}
\usepackage{bigdelim}
\usepackage{bm}
\usepackage{array}
\usepackage{longtable}
\usepackage{lscape}
\usepackage{float}
\usepackage{rotating}
\usepackage{multirow}
\usepackage{setspace}

\makeatletter

\newcommand{\Rmnum}[1]{\expandafter\@slowromancap\romannumeral #1@}
\makeatother

\begin{document}

\preprint{APS/123-QED}

\title{Gapless and Massive 1D Singlet Dispersion Channel in Infinite Spin-1/2 Ladders
---Infinite Quasi-1D Entanglement Perturbation Theory for Excitation\\
}
\author{Lihua Wang$^{1}$}
\email{wanglihua94@tsinghua.org.cn}
\author{Aslam Parvej$^{1}$}
\email{aslam18theo@unist.ac.kr}
\author{D. ChangMo Yang$^{1}$}
\author{Kwang S. Kim$^{1}$}
\email{kim@unist.ac.kr}
\affiliation{$^1$
Dept. of Chemistry, School of Natural Science, Center for Superfunctional Materials, Ulsan National Institute of Science and Technology (UNIST), Ulsan 44919, Republic of Korea}

\date{\today}

\begin{abstract} 
We solve for the elementary excitation in infinite quasi-1D quantum lattices by extending the recently developed infinite quasi-1D entanglement perturbation theory. The wave function of an excited state is variationally determined by optimizing superposition of cluster operation, each of which is composed of simultaneous on-site operation inside a block of lattice sites, on the ground state in a form of plane wave. The excitation energy with respect to the wave number gives the spectra for an elementary excitation. Our method is artificial broadening free and is adaptive for various quasi-particle pictures. Using the triplet spectrum, the application to $\infty$-by-$N$ antiferromagnetic spin-$\frac{1}{2}$ ladders for $N=2, 4, 6, 8$, and $10$ confirms a previous report that there is a quantum dimensional transition, namely, the lattice transits from quasi-1D to 2D at a finite critical value $N_c=10$. The massless triplet dispersion at $\left( \pi, \pi \right)$ sees a vanishing gap. Our results detect the anomaly at $\left(\pi,0\right)$ in the triplet spectrum, agreeing well with the inelastic neutron scattering measurement of a macroscopic sample. Surprisingly, our results also reveal a gapless and massive 1D singlet dispersion channel that is much lower than the triplet excitation. We note, however, the dimensional transition is determined by the massless triplet dispersion.      
\end{abstract}

\maketitle

An accurate calculation of the excitation energy (EE) of an infinite many-body system is faced with numerous challenges. On the one hand, the collective quantum fluctuation typically gives a finite difference between the ground state (GS) and excited state (ES) energies (both are infinite) for an infinite system. The commonly used strategy of calculating the energy difference per particle fails. A recent work\cite{Vanderstraeten2019} which embeds a single-site excitation tensor as a plane wave in an infinite projected entangled-pair state(iPEPS) represents a promising attempt. Yet, the progress along this line requires larger iPEPS bond index to provide convergence. Exploring the effect of non-local excitation operator is also necessary, as our work will show. As both analytical and numerical studies indicate that the physics in an infinite system is intrinsically different from that of a finite one\cite{Anderson1972}, precise simulations are desirable at regimes beyond which the system monotonically approaches the thermodynamic limit. For instance, the nonlinear sigma model\cite{Chakravarty1996,Sierra1996} previously predicted that the Heisenberg antiferromagnet (HAF) on an $\infty$-by-$N$ square lattice at zero temperature is ordered only when $N=\infty$. A recent numerical work\cite{Wang2019} showed that such a lattice is ordered when $N\ge 10$. Namely, the system undergoes a quantum dimensional transition from quasi-1D to 2D. 

On the other hand, it has been shown that various quasi-particle pictures may come into play at different excitation wavelengths. One example is provided by a 2D HAF on square lattices studied in this work. According to the linear spin-wave theory (LSWT)\cite{Hamer1992}, excitation is dominantly spin-1 Bosonic magnons and spin-0 weakly interacting pairs of magnons. It predicts that the magnon energy is maximal and constant along the line $q =\left(\pi ,0\right)$ to $\left(\pi/2,\pi/2\right)$. However, the recent inelastic neutron scattering (INS) measurements on Cu(DCOO)$_2\cdot$4D$_2$O (CFTD) \cite{Ronnow2001}, K$_2$V$_3$O$_8$\cite{Lumsden2006}, and La$_2$CuO$_4$\cite{Headings2010} show evidence which deviates from LSWT for the high-energy/short-wavelength dynamics\cite{Ronnow2001,Christensen2007}. At the momentum $ q = \left(\pi ,0\right)$ the spectrum envelope is suppressed compared with that at $\left(\pi/2,\pi/2\right)$, hereafter called an anomaly. A possible interpretation is that the states constituting the continuum correspond to different pairs of fractional excitation. Many works made progress to explore this anomaly. Series expansion with longer series\cite{Zheng2005} finds a $9.4\%$ anomaly considerably larger than the third-order spin-wave theory (SWT) prediction\cite{Hamer1992}. The modified-SWT\cite{Yamamoto2019} by including higher-order spin exchange couplings shows the raising of energy at $\left(\pi,0\right)$ with respect to that at $\left(\pi/2,\pi/2\right)$. Meanwhile, the dynamical structure factors along a path of highly symmetric points in Brillouin zone were calculated both by the density matrix renormalization group (DMRG)\cite{Verresen2018} and by quantum Monte Carlo (QMC)\cite{Shao2017}. But the numerical findings are also conflicting. Various calculations have pointed to a significant suppression of the magnon energy and an anomalously large continuum of excitation around $q = \left(\pi,0\right)$ \cite{Singh1995,Sandvik2001,Powalski2015,Powalski2018,Piazza2014}. Meanwhile, some other QMC results\cite{Reger1988,Runge1992,Sandvik1997,Sandvik2010,Jiang2011} are in good agreement with LSWT. Note that all the mentioned simulations were carried on finite HAF lattices.

In this work, we extend the infinite quasi-1D entanglement perturbation theory (iqEPT)\cite{Wang2015,Wang2019} to iqEPT for elementary excitation (iqEPT-e) to determine both the ES and the EE for an $\infty$-by-$N$ square lattice. We emphasize that our method is free of artificial broadening free, and so results in a sharp spectrum. Moreover, it is quasi-particle adaptive. As explained in detail later, the so-called cluster operator deployed in our method is composed of simultaneous on-site operation inside a block of lattice sites. When its size is larger than $1 \times 1$, varying it can detect either one-magnon or two-spinon states when the restriction $S^z_{total}=1$ is applied. $S^z_{total}$ counts the spin flipping in this cluster operator. Meanwhile, such cluster operators with $S^z_{total}=0$ may detect the anti-bounding two-magnon state. As a result, we not only clarify the anomaly on debate, but also discover a surprising gapless and massive 1D singlet dispersion channel that is much lower than the triplet excitation. We manifest the roles played by both singlet and triplet excitations in the previously reported quantum dimensional transition for such a model\cite{Wang2019}. In the rest of this letter, we interweave two lines of reasoning to achieve the extension of iqEPT to iqEPT-e. The first is essentials of iqEPT itself whose main character is to convert the $N$ sites in a rung of an $\infty$-by-$N$ lattice to an effective site of an infinite chain and then to build and solve for a matrix product state (MPS)\cite{Oestlund1995,Verstraete2004a,Chung2006} on it. In this way both the number of significant diagonalized reduced density matrix elements and the entanglement entropy for an effective site are shown to saturate when $N$ increases, making iqEPT efficient for large $N$'s. The second line follows a previous work\cite{Chung2009} using Feynman's idea\cite{Feynman1954} to precisely retrieve ES by projecting excitation operators onto GS as the virtual vacuum.

The model Hamiltonian is
\begin{equation}
\label{eq:hamiltonian}
H=J\sum_{\lceil \left(i,j\right),\left(i',j'\right)\rceil}{{\vec S}_{\left(i,j\right)}\cdot {\vec S}_{\left(i',j'\right)}}, 
\end{equation} 
where ${\vec S}_{\left(i,j\right)}$ is the spin vector operator on the ${\left(i,j\right)}^{\text{th}}$  lattice site with $\it{i}$  running from $-\infty$ to $\infty$ in the longitudinal direction (LD) and $\it{j}$ running from $1$ to $N$ in the transverse (rung) direction (TD). $\lceil\cdots\rceil$ sums over the nearest neighboring sites. $J$ is the spin-spin coupling integral and is normalized to $1$ hereafter. The periodic boundary condition (PBC) is assumed in both directions.

As mentioned, the GS $\mid g\rangle$ obtained by iqEPT is expressed as an MPS 
\begin{equation}
\label{eq:mps0}
\mid g\rangle=\sum_{\cdots r^{i-1}r^i\cdots}{tr\left(\cdots \xi_{r^{i-1}}\cdot \xi_{r^i}\cdots\right)\cdots\mid \phi_{r^{i-1}}^{i-1}\rangle\mid \phi_{r^{i}}^{i}\rangle\cdots} 
\end{equation}
where $r^i$ runs from $1$ to $L^N$, with $i$ being the coordinate of $i^{\it{th}}$ effective site and $L$ being the local space rank of an original lattice site ($2$ for spin-$\frac{1}{2}$). $\phi_{r^{i}}^{i}$ is a local state vector and $\xi_{r^i}$ is a matrix assigned to such a vector. We start with a trial wave function $\mid \widetilde{g}^0_P\rangle$ for a given MPS rank $P$ and iteratively optimize $\mid \widetilde{g}_P\rangle$ over each $\xi_{r^i}$ by solving a series of generalized eigenvalue equations (GEE) generated from $\frac{\partial {\langle \widetilde{g} \mid H\mid \widetilde{g}\rangle}/{\langle \widetilde{g} \mid  \widetilde{g}\rangle}}{\partial \xi_{r^i}}=0$, until $\mid \widetilde{g}_P\rangle$ converges to $\mid g_P\rangle$ with a set of MPS tensors $\{\xi_{r^i}\}_P$. We repeat this process for a slightly enlarged MPS rank $P+\Delta P$ with small new elements appended to each tensor in $\{\xi_{r^i}\}_P$ as the perturbation to make $\{\xi_{r^i}\}_{P+\Delta P}$ and hence $\mid \widetilde{g}^0_{P+\Delta P}\rangle$. Eventually, $\mid g_P\rangle$ converges to the true solution $\mid g\rangle$.   

Meanwhile, we extend the cluster-operator representation from ref \cite{Chung2009} for the solution to ES. We begin by associating an ES with a 2D wave-number $\left(k_x,k_y\right)$ as  
\begin{equation}
\label{eq:excitation_wavefunction}
\mid \Phi_{k_x k_y}\rangle\equiv\sum_{\substack{m=-\infty,\infty\\n=1,N}}{e^{\bar{i}\left(k_x m+k_y n\right)}\Theta^{\dagger}_{mn}\mid g\rangle}.
\end{equation}
The cluster operator $\Theta^{\dagger}_{mn}$ on $\mid g \rangle$ is defined as
\begin{align}
\label{eq:superposition}
\Theta^{\dagger}_{mn}\equiv c_l\Xi^{\dagger l}_{mn}.
\end{align}
This is the superposition of a simultaneous operation (direct product) of $a\times b$ operators on a $a\times b$ rectangular block of lattice sites whose anchor site is $\left(m,n\right)$. Each of such simultaneous operation is defined as
\begin{align}
\label{eq:cluster_operator}
\Xi^{\dagger l}_{mn}\equiv {\prod_{\substack{i=0,a-1\\j=0,b-1}}{\hat{o}^{\dagger l}_{m+i,n+j}}}.
\end{align}
Evidently, the conjugate of this simultaneous operation is the direct product of each individual conjugate operator. If each site allows for $M$ possible operations, the complete configuration space has a rank $M^{ab}$. Furthermore, one may define its subspace by specifying a quantum number, for instance, the total z-component $S^z_{total}$ of spins in a $a\times b$ site block. $l$ runs from $1$ to the rank $r$ of such a subspace. 

The EE between the ES energy and the GS energy $\epsilon_{g}=\langle g\mid H \mid g\rangle$ is 
\begin{align}
\label{eq:energy_gap}
&E^g_{k_x k_y} \notag\\ 
=&{\frac{\sum_{ij}\sum_{\substack{m=-\infty,\infty\\n=1,N}}{e^{-\bar{i}\phi_{k_xk_y}^{mnij}}\langle g\mid \Theta_{mn} \left[H,\Theta_{ij}^{\dagger}\right]\mid g\rangle}}{\sum_{i'j'}\sum_{\substack{m=-\infty,\infty\\n=1,N}}{e^{-\bar{i}\phi_{k_xk_y}^{mni'j'}}\langle g\mid \Theta_{mn}  \Theta_{i'j'}^{\dagger}\mid g\rangle}}}
\end{align}
where $\phi_{k_xk_y}^{mnij}=k_x\left(m-i\right)+k_y\left(n-j\right)$. We used the translational symmetry. The x-coordinate $i$ ($i'$) and y-coordinate $j$ ($j'$) run over the lattice site inside a unit cell (owing to the bipartite nature of the HAF, they run from $1$ to $2$ in this study). We substitute equations \eqref{eq:superposition} and \eqref{eq:cluster_operator} into equation \eqref{eq:energy_gap} to obtain
\begin{align}
\label{eq:energy_gap_matrix}
E^g_{k_x k_y}=\frac{c_u \Upsilon_{uv}^{k_xk_y} c_v}{c_{u'} \Lambda_{u'v'}^{k_xk_y} c_{v'}}=\frac{C^T \Upsilon^{k_xk_y} C}{C^T \Lambda^{k_xk_y} C},
\end{align} 
where we have converted the superposition coefficient $\{c_l\}$ into a column matrix (or a vector) $C$ and have defined two matrices $\Upsilon^{k_xk_y}$ and $\Lambda^{k_xk_y}$ whose elements are given as
\begin{align}
\label{eq:energy_gap_matrix0}
&\Upsilon_{uv}^{k_xk_y} =\sum_{ij}\sum_{\substack{m=-\infty,\infty\\n=1,N}}{e^{-\bar{i}\phi_{k_xk_y}^{mnij}}\langle g\mid \Xi^u_{mn} \left[H,\Xi_{ij}^{\dagger v}\right]\mid g\rangle}\notag\\
&\Lambda_{uv}^{k_xk_y} =\sum_{ij}\sum_{\substack{m=-\infty,\infty\\n=1,N}}{e^{-\bar{i}\phi_{k_xk_y}^{mnij}}\langle g\mid \Xi^u_{mn} \Xi_{ij}^{\dagger v}\mid g\rangle}.
\end{align}
The variation of $E^g_{k_x k_y}$ with respect to $C$ leads to
\begin{align}
\label{eq:gee}
\Upsilon^{k_xk_y} C=E^g_{k_x k_y}\Lambda^{k_xk_y} C.
\end{align}
Thus, the optimal superposition coefficients $\bar{C}$ is obtained by solving the generalized eigenvalue equation \eqref{eq:gee} and is associated with the lowest excitation energy gap $\bar{E}^g_{k_x k_y}$. Finally, we substitute $\bar{C}$ into equations \eqref{eq:cluster_operator}, \eqref{eq:superposition}, and \eqref{eq:excitation_wavefunction} to get wave function of lowest excitation.

\begin{figure}
	\begin{center}
		\includegraphics[width=\columnwidth]{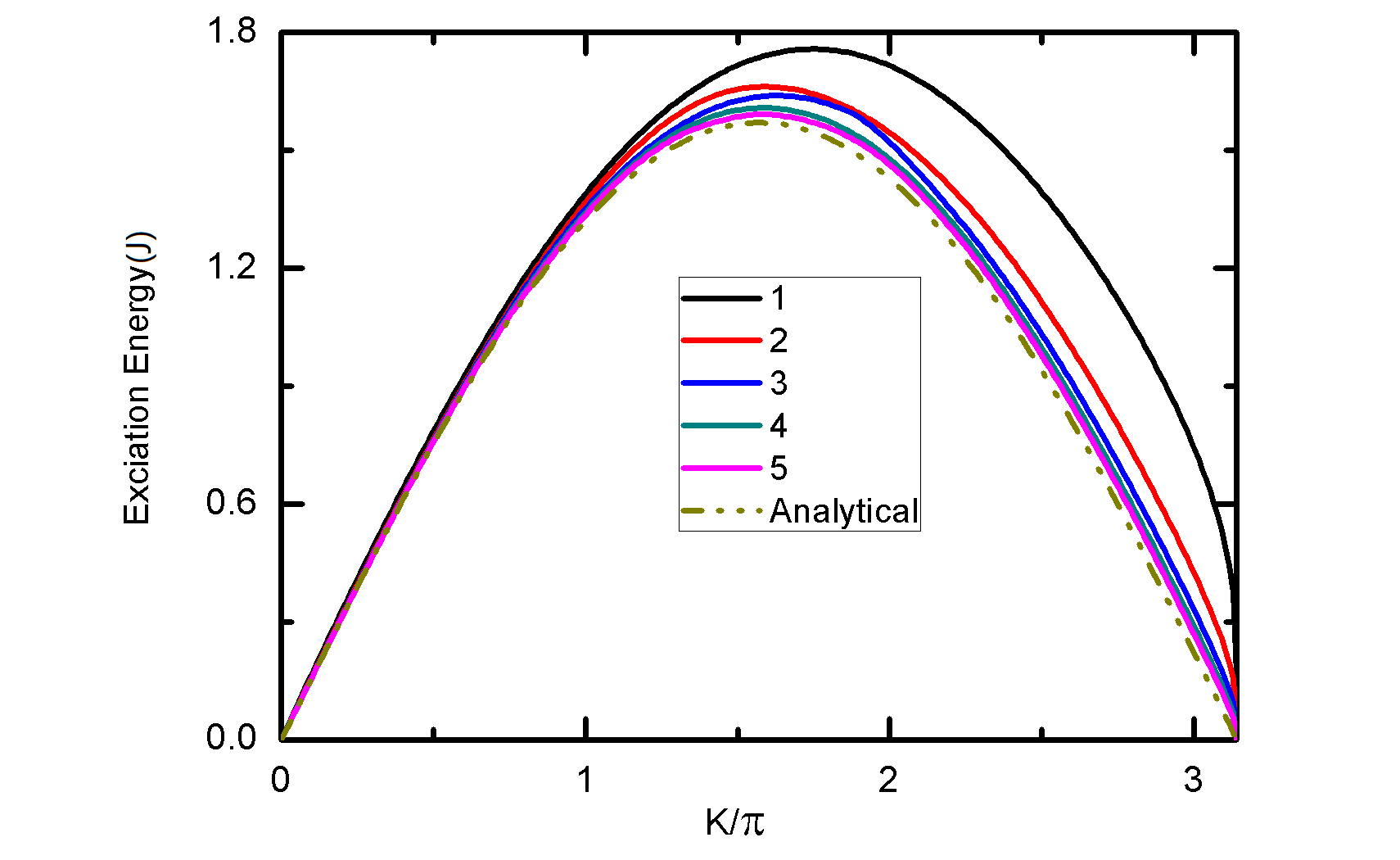}
		\caption{\label{fig:spinchain} Triplet spectrum for an infinite spin chain. From top to bottom, the solid curves for cluster operator size 1, 2, 3, 4, and 5 converge to the dashed curve for the exact Bethe ansatz result.}
	\end{center}
\end{figure} 

\begin{figure}
	\begin{center}
		\includegraphics[width=.9\columnwidth]{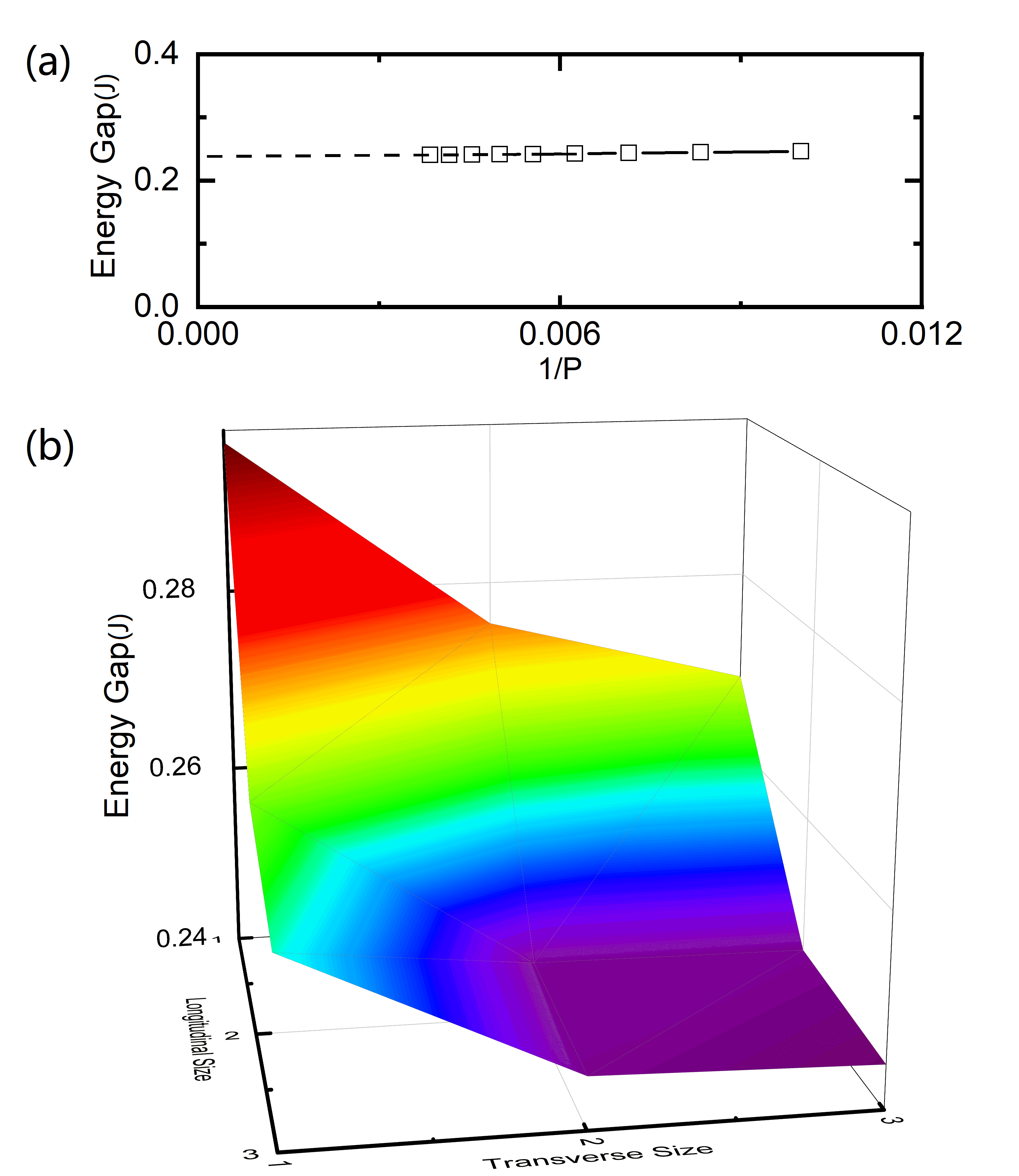}
		\caption{\label{fig:cluster_convergence} Convergence of energy gap at $\left(\pi,\pi\right)$ for a triplet excitation of an $\infty$-by-$4$ lattice. (a). The curve for cluster operator size $2\times 2$ is converged with $P$ starting from $P=100$. (b). Results for various operator sizes are given for $P=100$, by varying both longitudinal and transverse sizes from $1$ to $3$.}
	\end{center}
\end{figure} 

There are few technical notes.

1. The commutation in equation \eqref{eq:energy_gap_matrix0} can be explicitly simplified, as the supplemental information shows.

2. The cluster operator may be constrained with the value of $S^z_{total}$ which counts the spin flipping in this cluster operation, $0$ for singlet and $1$ for triplet excitations.
 
3. The simultaneous operation $\Xi^{\dagger l}_{mn}$ on an $a\times b$ rectangular block of lattice sites needs to be re-associated according to the scheme used by iqEPT to convert each rung of width $N$ into an effective site. We first perform the direct product inside a rung which crosses with a cluster operator by inserting identity operator $\it{I}$ on the un-operated original lattice site to form an auxiliary operator. Finally, we obtain $\Xi^{\dagger l}_{mn}$ by performing the direct product among all auxiliary operators.      

4. When choosing a complete subspace, one should avoid over-completeness. For instance, for a triplet excitation, the complete subspace ranks for cluster sizes $1$, $2$, $3$, and $4$ should be $1$, $4$, $14$ (rather than $15$), and $52$ (rather than $56$), respectively. For example, for the cluster size $3$, the configuration $I\otimes I\otimes S^+$ is not independent of $S^+\otimes I\otimes I$ in a background composed of one identity operator on each original lattice site.  

5. There are three scenarios for the correlation function $\langle \Theta_{mn} \Theta_{ij}^\dagger\rangle$ where $m$ and $i$ are the LD coordinates. $n$ and $j$ are the TD coordinates. 

a. Exponential decay. It is safe to truncate the summation in equations \eqref{eq:energy_gap} and \eqref{eq:energy_gap_matrix0} as $\sum_{m=-1000, 1000}$. 

b. Ordered. It is safe to set the summation in equations \eqref{eq:energy_gap} and \eqref{eq:energy_gap_matrix0} as $\sum_{m=-m_0, m_0}$ in that the correlation function varies no more beyond separation $m_0$. In the case of an $\infty$-by-$10$ lattice studied in this paper, $m_0=1000$ is adequate.

c. Quasi-long-range order. An infinite spin chain falls in this category. Our results show that the summation range should be larger for larger cluster sizes. For the cluster size $5$, it is safe to set the summation in equations \eqref{eq:energy_gap} and \eqref{eq:energy_gap_matrix0} as $\sum_{m=-20000, 20000}$. It requires an extremely accurate simulation of $\mid g\rangle$. 

\begin{figure}
	\begin{center}
		\includegraphics[width=.9\columnwidth]{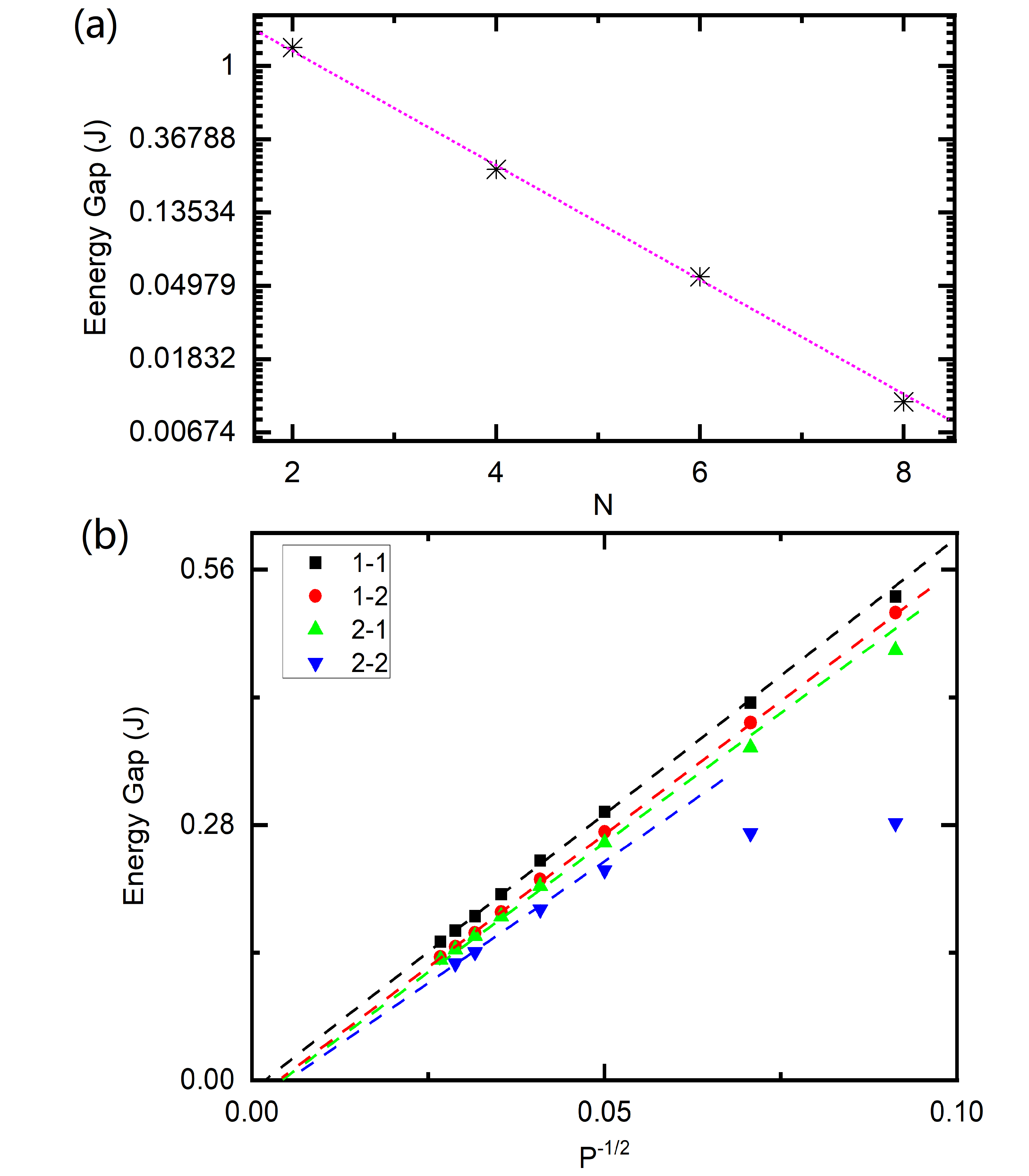}
		\caption{\label{fig:dimension-transition} Energy gap at $\left(\pi,\pi\right)$ for a triplet excitation closes up at a critical ladder width $N_c=10$. (a) The linear log-log relation between the energy gap and $N$ confirms the nonlinear Sigma model's prediction of the exponentially closing up gap fully for $N=2$, $4$, and $6$; partially for $N=8$. (b) When the MPS rank $P\rightarrow \infty$, the computed energy gap for $N=10$ will drop to $0$ at finite MPS ranks.}
	\end{center}
\end{figure} 

First, we benchmark our algorithm on an infinite spin chain. We obtained the GS energy per spin $-0.44314717$ at the largest MPS rank $600$ deployed in iqEPT. In exact agreement with Bethe ansatz GS energy\cite{Bethe1931}, this allows for precise computation of spin-spin correlations for separations over $10000$ sites. The correlations reproduced results in ref \cite{Wang2012}, both of which agree with the prediction by the conformal field theory\cite{Affleck1998a} that there is a logarithmic correction multiplicative to the power-law decay predicted by the bosonization theory\cite{Luther1974}. As Fig.\ref{fig:spinchain} shows, the triplet excitation results for cluster sizes $1$, $2$, $3$, $4$, and $5$ have converged to the exact Bethe ansatz triplet (lowest) excitation spectra. Note that our results resemble the spectra obtained in refs\cite{Chung2009,Yamamoto1995} but do not require finite-size scaling.

Next, we discuss both triplet and singlet excitations for the target model, HAF on $\infty$-by-$N$ square lattices, respectively. Many of their characteristics with respect to the parameter settings such as the MPS rank and the cluster operator size are transferable to each other. Thus, we discuss the triplet excitation first.
 
We show the convergence behavior by choosing the energy gap at $\left(\pi,\pi\right)$ for $N=4$ as an example. Fig.\ref{fig:cluster_convergence} (a) shows that the result for a relatively large $2\times 2$ cluster operator is converged with $P\ge 100$. Using data obtained at $P=100$, Fig.\ref{fig:cluster_convergence} (b) shows another convergence with respect to the cluster operator sizes along LD and TD. It is safe to say cluster operator size $2\times 2$ is adequate. Hereafter, all results are given for the $2\times 2$ cluster operator unless specified otherwise. Our results show that a yet to made convergence with the iPEPS bond index in ref\cite{Vanderstraeten2019} is, generally speaking, equivalent to a convergence for the cluster size $1\times 1$ here.

Further verification for the previously reported quantum dimensional transition\cite{Wang2019} is made with a more direct signal of energy gap itself, as shown in Fig.\ref{fig:dimension-transition}. Indeed, the gap at $\left(\pi,\pi\right)$ for a triplet excitation vanishes for $N\ge 10$.

\begin{figure}
	\begin{center}
		\includegraphics[width=.9\columnwidth]{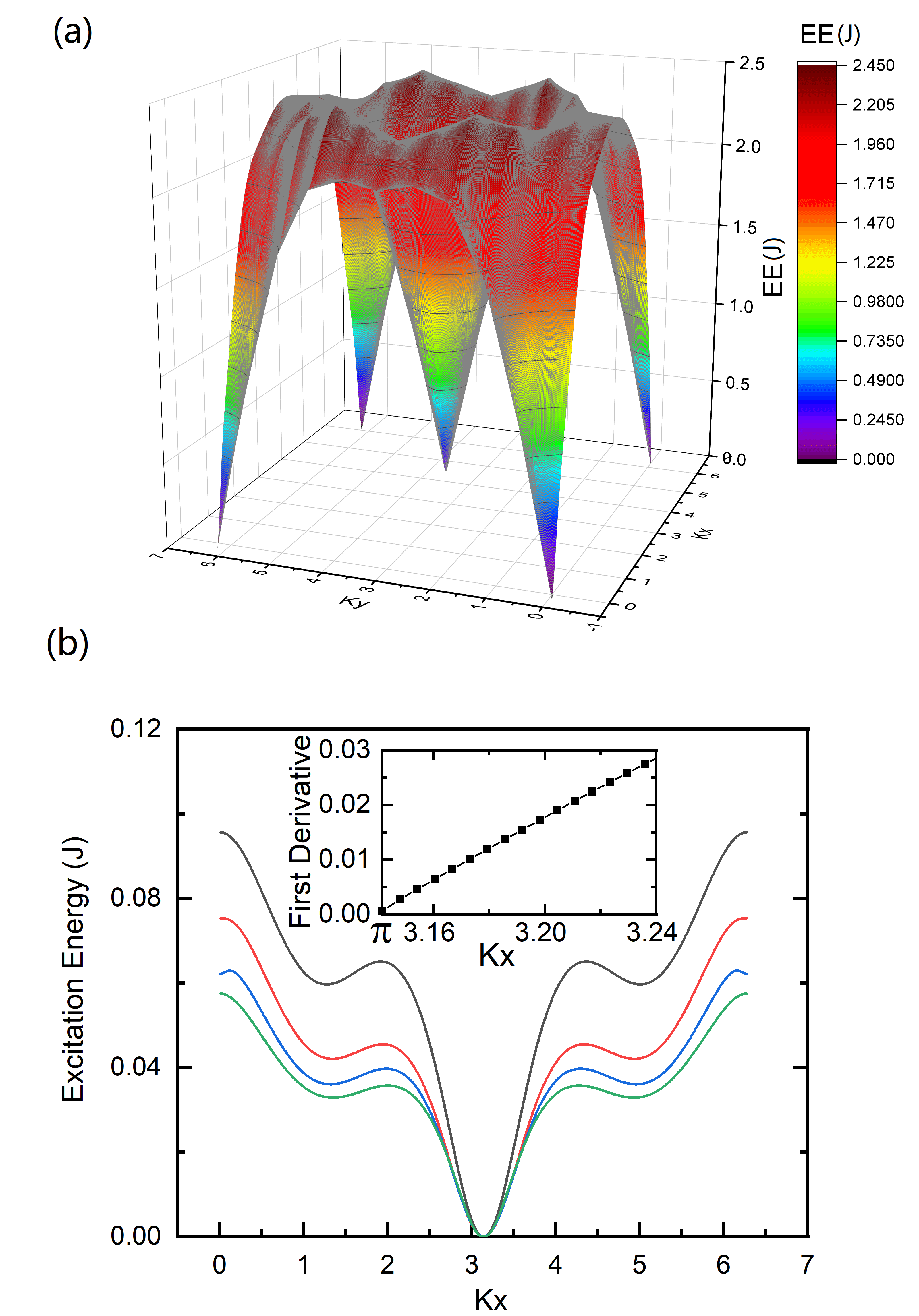}
		\caption{\label{fig:spectra} Spectra for an $\infty$-by-$10$ lattice.}
	\end{center}
\end{figure}

Fig.\ref{fig:spectra} (a) shows the triplet spectrum for an $\infty$-by-$10$ lattice. It restores the square lattice geometric symmetry with those Goldstone-mode k-points. Namely, at those high-symmetry points, the dispersion is gapless and massless so as to allow a spontaneous symmetry breaking\cite{Goldstone1962}. Noticeably, there is a normaly at $\left(\pi,0\right)$ compared with $\left(\pi/2,\pi/2\right)$. Limited by commensuration, we choose the $N=8$ spectrum ($N=8$ and $N=10$ yield similar high-energy spectra) for the quantitative comparison of the anomaly with experiments. EE is $2.34228J$ at $\left(\pi,0\right)$ and $2.46696J$ at $\left(\pi/2,\pi/2\right)$. There is pronounced anomaly of $5.3\%$ which is close to the latest experimental observation of $7\%$ anomaly for a macroscopic CFTD HAF sample\cite{Wan2020}. 

Lastly, our results show that there is a 1D channel of $k_y=0$ or $2\pi$ where the singlet dispersion is much lower than the triplet dispersion except the small vicinity around $k_x=0$ or $2\pi$. More profound is that at $\left(\pi,0\right)$ the singlet dispersion is gapless (Fig.\ref{fig:spectra} (b)) and massive (Inset of Fig.\ref{fig:spectra} (b)). In contrast, the triplet dispersion sees a dome at this k-point (Fig.\ref{fig:spectra} (a)). Note that the overall excitation energy scale of singlet is more than $10$ times smaller than that of triplet in this channel. We attribute this to the anti-bounding between two magnons with opposite spins. It is challenging for experiments performed at any finite temperature to reveal such a delicate singlet excitation due to its small energy scale.  

In conclusion, we developed a new method, iqEPT-e, to simulate the elementary excitation in a truly infinite quantum lattice. It is variational and free of artificial spectral broadening. The observation of anomaly at $\left(\pi,0\right)$ for the triplet excitation of HAF on $\infty$-by-$N$ lattices quantitatively agrees well with the latest INS measurement on a macroscopic CFTD sample. Moreover, our method is quasi-particle adaptive so as to capture a special 1D dispersion channel where the singlet excitation has much lower energy than the triplet excitation. Since the singlet 1D channel is massive, despite being gapless, the previously reported quantum dimensional transition is determined by the massless triplet excitation. It happens when the triplet excitation gap at $\left(\pi,\pi\right)$ vanishes for an $\infty$-by-$N (\ge 10)$ spin ladder.    

\hspace{11cm}

\begin{acknowledgments}
	This work was supported by NRF (National Honor Scientist Program 2010-0020414) and KISTI (KSC-2019-C3-0081). 
	
	L.W. developed the method, programmed the codes and wrote the paper. A.P. contributed to coding and writing. D.C.Y. commented the paper. K.S.K. supervised the work.
\end{acknowledgments}

\bibliography{iqEPTe}

\end{document}